\documentclass[aip,rsi,reprint,secnumarabic,amssymb,amsmath,nobibnotes,showpacs]{revtex4-1}

\usepackage{graphics} \usepackage{graphicx} \usepackage{braket} \usepackage{float} \usepackage{tabularx} 
\usepackage[hyperindex,breaklinks]{hyperref}
\usepackage[usenames,dvipsnames]{color}

\begin{document}

\title{A 3D-printed alkali metal dispenser}

\author{E. B. Norrgard}

\email[Electronic address: ]{eric.norrgard@nist.gov}
\affiliation{Joint Quantum Institute, National Institute of Standards and Technology and University of Maryland, Gaithersburg, Maryland 20899, USA}

\author{D. S. Barker}
\affiliation{Sensor Sciences Division, National Institute of Standards and Technology, Gaithersburg, MD 20899, USA}
\author{J. A. Fedchak}
\affiliation{Sensor Sciences Division, National Institute of Standards and Technology, Gaithersburg, MD 20899, USA}
\author{N. Klimov}
\affiliation{Sensor Sciences Division, National Institute of Standards and Technology, Gaithersburg, MD 20899, USA}
\author{J. Scherschligt}
\affiliation{Sensor Sciences Division, National Institute of Standards and Technology, Gaithersburg, MD 20899, USA}
\author{S. Eckel}
\affiliation{Sensor Sciences Division, National Institute of Standards and Technology, Gaithersburg, MD 20899, USA}
\begin{abstract} We demonstrate and characterize a source of Li atoms made from direct metal laser sintered titanium. The source's outgassing rate is measured to be $5 \,(2)\cdot 10^{-7}$\,$\rm{Pa}~ \rm{L}~ \rm{s}^{-1}$ at a temperature $T=330\,^\circ$C, which optimizes the number of atoms loaded into a magneto-optical trap.  The source loads $\approx 10^7$ $^7$Li atoms in the trap in  $\approx 1$\,s. The loaded source weighs 700\,mg and is suitable for a number of deployable sensors based on cold atoms.  \end{abstract}


\maketitle

Numerous emerging quantum technologies are being adapted  from laboratory-scale experiments to deployable real-world sensors and space-based missions \cite{Wang2007b,Ladouceur2009, Hogan2011,Muntinga2013}.
Cold atoms are at the heart of many of these applications, including interferometers \cite{Hogan2011} and  ultra-precise atomic clocks \cite{Beloy2014, Marti2017}.  Recently, we have begun a program to develop a device based on trapped cold atoms which is simultaneously a primary standard and an absolute sensor of vacuum \cite{Scherschligt2017,Jousten2017}.

Translating these cold atom-based technologies into deployable sensors requires an atomic source which is scaleable, lightweight, and suitable for ultra-high vacuum (UHV, $<10^{-6}$\,Pa).  Miniature vapor cells are well suited for sensors using atoms which have substantial vapor pressure at room temperature (such as Rb) \cite{Perez2009, Mhaskar2012}.  However, many applications use  atoms which must be heated to several hundred \mbox{degrees Celsius} to produce a substantial vapor pressure, including Li, Sr, and Yb \cite{Ladouceur2009, Beloy2014, Marti2017}.

Several groups have reported direct loading of a magneto-optical trap (MOT) from effusive sources \cite{Moore2005,Muhammad2008,Ladouceur2009,Scherer2012} such as ovens or alkali metal dispensers (AMDs).  Some commercial AMDs are known to outgas at levels not suitable for ultra-high vacuum \cite{Muhammad2008}  when operating.  This can be mitigated by including a reducing agent, but these typically limit the alkali yield to $\approx$\,10\,mg.
Alternatively, the small features and thin walls (for high electrical resistance) of an AMD makes additive manufacturing an attractive construction technique. Direct metal laser sintered (DMLS) titanium  has recently been demonstrated to be low outgassing and suitable for in-vacuum components \cite{Gans2014} and  vacuum flanges \cite{Vovrosh2017}.

In this note, we report on a lithium AMD made from DMLS grade 5 titanium (6\,\% Al, 4\,\% V by weight).  The AMD holds $\approx 100$\,mg of Li and is used to directly load a MOT suitable for use in cold atom experiments.  The  measured outgassing rate is \mbox{$5\,(2)\cdot 10^{-7}$\,$\rm{Pa}~ \rm{L}~ \rm{s}^{-1}$} for optimal MOT loading conditions (uncertainties in this paper are the uncorrelated combination of $1\sigma$ statistical and systematic uncertainties), limited by outgassing from the loaded lithium metal.


\begin{figure}[b]
  \centering
  \includegraphics[width=\columnwidth]{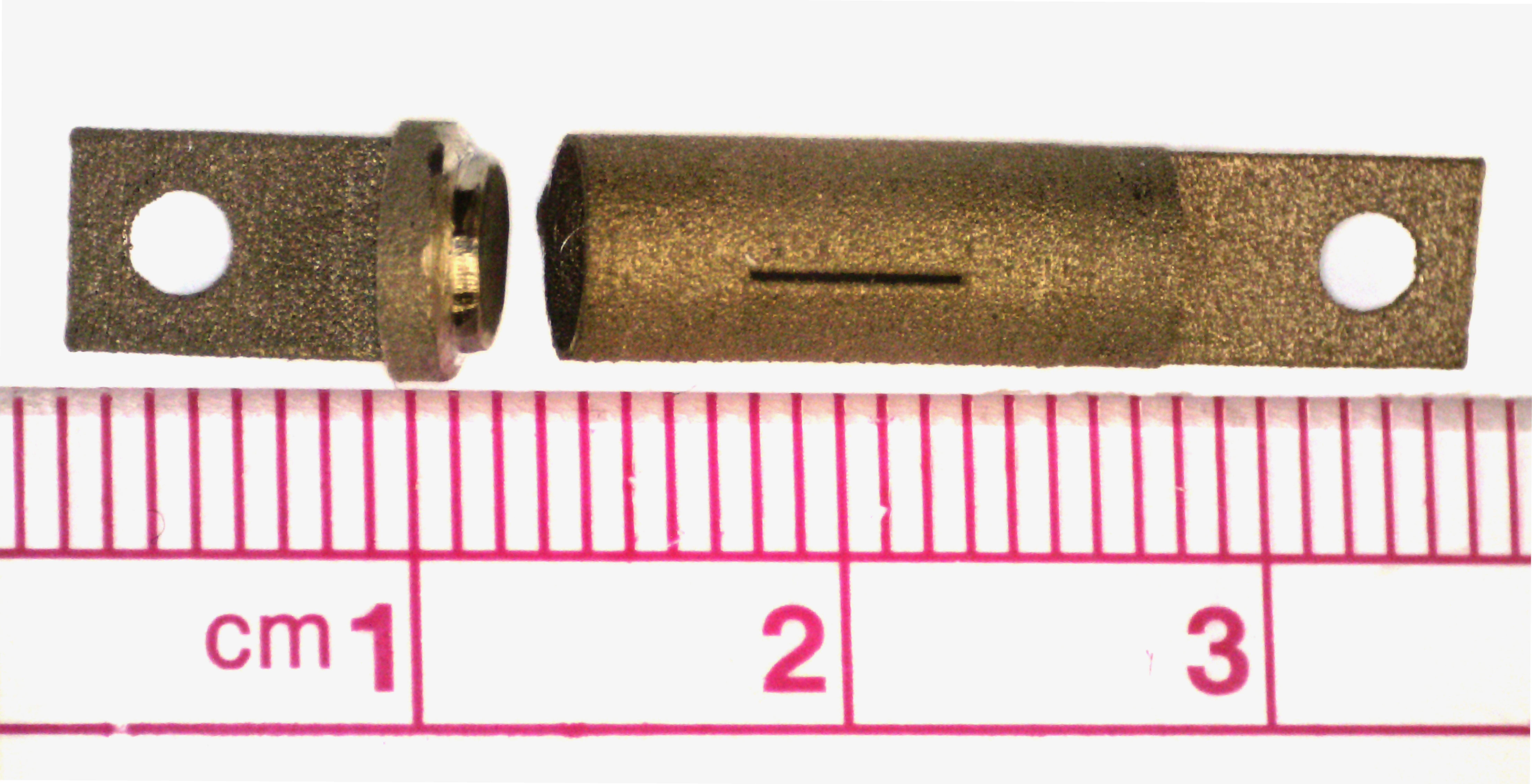}
  \caption{(Color online) Photograph of the 3D-printed titanium cap (left) and tube (right).}\label{fig:photo}
\end{figure}

Figure \ref{fig:photo} shows the 3D-printed AMD, which consists of two parts: a tube (5.1 mm diameter, 13.8 mm long, 0.13 mm wall) and a plug (1.2 mm thick), which snugly fits into the open end of the tube.  Both the tube and plug have 0.25 mm thick, 5.1 mm wide tabs with a 2.4 mm diameter clearance hole for mounting and electrical connections.  A 5.1 mm long, 0.25 mm wide slit in the tube roughly directs the exiting alkali atoms orthogonal to the plane of the tabs.  The combined tube and plug have a measured mass of 584\,(2)\,mg and designed total surface area (not including surface roughness) of 6.8\,cm$^2$.

The AMD is loaded with metallic lithium and inserted into the  vacuum chamber while under an argon-purged atmosphere. We loaded the AMD with seven pellets of natural isotopic abundance Li, which we estimate to total 100\,mg.  The plug fits tightly into the tube to secure the Li. The AMD is mounted on oxygen-free, high-conductivity Cu electrical feedthroughs 35\,mm from the center of the MOT.

The AMD is resistively heated by a current of typically 10\,A to 15\,A.  Upon initial warm up, the resistance of the AMD dropped by nearly a factor of two, presumably as the (relatively conductive) Li melted and came into better electrical contact with the tube.  We therefore characterize the AMD in terms of the power $P$ dissipated across the AMD.  With subsequent operations, we have noticed further decreases in the AMD resistance at the 1\,\% level, and noticed a small accumulation of Li outside the slit of the tube.   A thin Ni mesh inside the tube could be used to wick the Li in a future design\cite{Gunton2013}.

A laser beam counterpropagating to the Li beam allows the temperature $T$ of the Li beam to be determined spectroscopically. We monitor the laser induced fluorescence collected on a charge-coupled device (CCD) camera from a spatial region along the laser beam. The frequency-dependent fluorescence is fit to a Maxwell-Boltzmann distribution for a 1D beam.
 The temperature was measured for a number of powers and fit to an empirical model containing conductive ($P \propto T-T_{\rm{r}}$) and radiative ($P \propto T^{4}-T_{\rm{r}}^4$) terms, where $T_{\rm{r}}=20\,(1)\,^\circ$C is the measured room temperature.  In the experiments discussed below, this fit was used to determine the Li beam temperatures from the measured power dissipated in the AMD.

All stainless steel components of the vacuum chamber were baked at 450\,$^\circ$C for 21 days prior to assembly to reduce hydrogen outgassing \cite{Santeler1992,Sefa2017}.  All vacuum components, including the 3D-printed titanium AMD, were cleaned in alkaline detergent, followed by acetone, then ethanol.  No cleaning procedure was performed on the Li beyond selecting pellets with minimal black nitride coating.

\begin{figure}[b]
  \centering
  \includegraphics[width=\columnwidth]{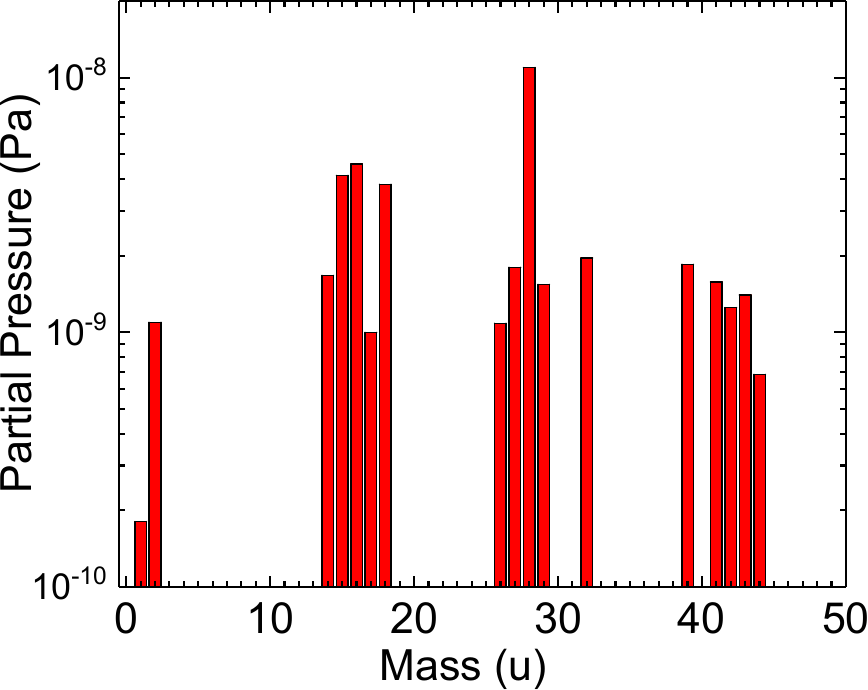}
  \caption{(Color online) Background-subtracted (i.e. source on minus source off) mass spectrum of gas composition with \mbox{$T=330\,^\circ$C}.  The total pressure including background is \mbox{$3.3\,(6) \cdot 10^{-7}$\,Pa}.}\label{fig:histogram}
\end{figure}

The chamber was pumped by a 50\,L\,s$^{-1}$ ion pump and monitored by a mass spectrometer (Granville-Phillips Vacuum Quality Monitor, with an ion gauge mounted nearby measuring total pressure \cite{NISTDisclaimer}).  The AMD was degassed for 3 days by dissipating $P$\,=\,2.0\,W ($T \approx 240\,^\circ$C).  We did not otherwise bake the vacuum chamber to remove water. Upon initial degassing, the pressure increased to $3\cdot 10^{-5}$\,Pa, dominated by water.  After degassing, the mass spectrometer  recorded a significant decrease in all gasses except for $m =28$ u.   Outgassing that is not a strong function of time is likely due to species that are chemically bound or are diffusing from bulk material. Because Li forms a nitride layer in the presence of air, we suspect the $m =28$ peak to be N$_2$ originating from the Li pellets.

Figure \ref{fig:histogram} shows the increase in background gas composition recorded on the mass spectrometer when operating the AMD under typical conditions ($P=3.0$\,W, $T\approx 330\,^\circ$C,  total pressure $3.3\,(6) \cdot 10^{-7}$\,Pa). The presence of a $m = 14$ u peak roughly $1/10$ the intensity of the $m = 28$ u peak is consistent with the cracking fraction of N$_2$ in most mass spectrometers.  The lack of a visible $m = 12$\,u peak indicates the possible contribution of CO to the $m = 28$ u peak is small. The magnitude of the observed $m = 44$ u peak indicates a negligible contribution to the $m = 28$ u peak from cracking of CO$_2$ into CO (28 u) and O (16 u).   In addition, the partial pressure of O$_2$ (32 u)  is observed to increase.  This too may originate from the Li pellets because Li forms hydroxides and carbonates upon exposure to air.

\begin{figure}[b]
  \centering
  \includegraphics[width=\columnwidth]{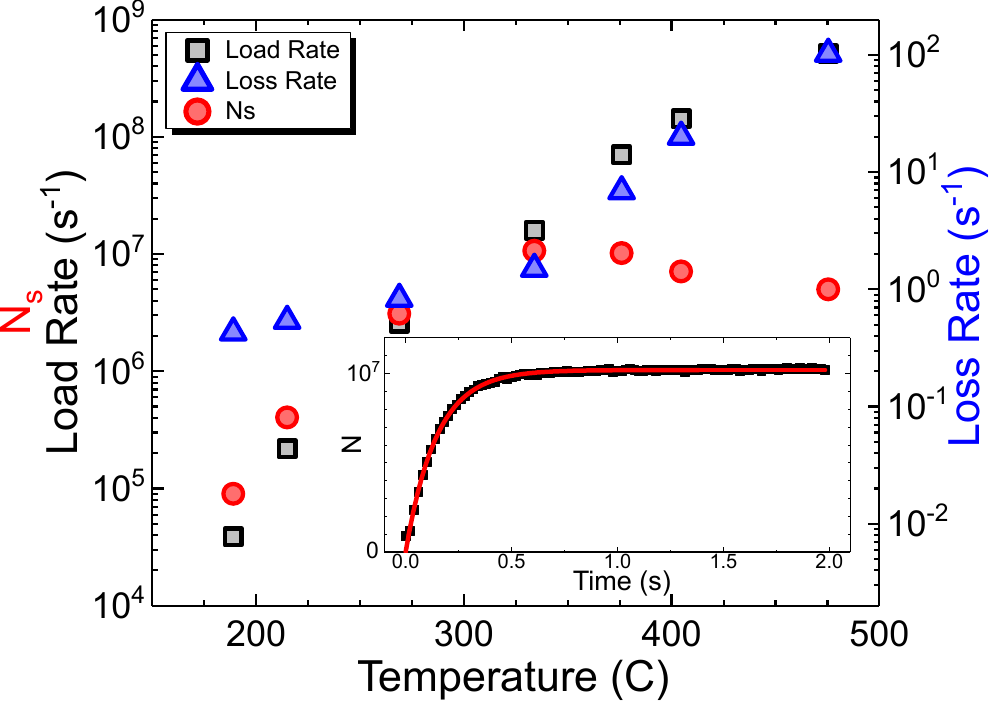}
  \caption{(Color online) Loading rate $R$ (black squares), steady-state atom number $N_{\rm{S}}$ (red circles), and one-body loss rate $\gamma$ (blue triangles) as a function of temperature $T$.  The inset shows a typical MOT loading curve with $T=330\,^\circ$C.  The red line is a fit to Eq.\,\ref{eq:Load}.}\label{fig:loading rate}
\end{figure}

We investigated the outgassing rate of the AMD by the throughput method, wherein the pressure $p$ in a vacuum chamber is determined by the total gas flow $q$ from all sources into the chamber and effective pumping speed $S$, $q=pS$.  The AMD was tested in a small stainless-steel vacuum chamber evacuated by a turbomolecular pump.   The effective pumping speed $S$ was limited by the conductance of the vacuum components between the AMD and the turbomolecular pump; we roughly estimate $S \approx  7\,\rm{L}\,\rm{s}^{-1}$ for N$_2$ at $T_{\rm{rm}}=20\,^\circ$C. The outgassing rate for the AMD $q_{\rm{AMD}}$ is determined by the measured pressure rise  $p_{\rm{AMD}}$ above the background when the source is turned on. We obtain an N$_2$ outgassing rate of $q_{\rm{AMD}}$\,=\,$5\,(2) \cdot 10^{-7}\,\rm{Pa}~ \rm{L}~ \rm{s}^{-1}$  for the source operating at $330\,^\circ$C.  Our estimate of $q_{\rm{AMD}}$  does not take into account possible outgassing of the chamber due to heating from the AMD, and therefore represents an upper limit.


The MOT consists of six independent, circularly polarized laser beams, detuned $-18$\,MHz from the $F=2\rightarrow F^\prime=3$ transition of the Li $D_2$ line.  Each beam has Gaussian rms width 3.6\,(2)\,mm and power 40\,(1)\,mW.  An electro-optic modulator adds 814\,MHz RF sidebands ($\approx 20$\,$\%$ in each of the  $\pm 1$ order sidebands) to the beams in order to drive the  $F=1\rightarrow F^\prime=2$ repump transition.  A quadrupole magnetic field with axial gradient $\frac{\partial B}{\partial z}$\,=\,3\,mT/cm is formed by two arrays of grade N52 permanent magnet bars held in 3D-printed acrylonitrile butadiene styrene mounts around the vacuum chamber.

The number of atoms $N$ in the MOT is determined by fluorescence imaging on a CCD camera.  We estimate the measured atom number to be accurate to within a factor of 2. The load dynamics of the MOT are well described by the differential equation
\begin{equation}\label{eq:loaddiffeq}
  \frac{dN}{dt} = R - \gamma N - \beta\int n^2 d^3x,
\end{equation}
Here, $n$ is the atomic density, $R$ is the trap loading rate, and $\gamma$ and $\beta$ are the one- and two-body loss rate coefficients, respectively.  Equation \ref{eq:loaddiffeq} has solution
\begin{eqnarray}
  N(t) &=& N_{\rm{S}} \frac{1-e^{-\gamma_0 t}}{1+\frac{N_{\rm{S}}^2 \beta}{V R}e^{-\gamma_0 t}},\label{eq:Load} \\
  \gamma_0 &=& \gamma \sqrt{1+\frac{4 \beta R}{V \gamma^2}},
\end{eqnarray}
where $N_{\rm{S}}$ is the steady-state atom number, and we  assume the MOT occupies a fixed volume $V$ determined by the fitted Gaussian widths on the CCD camera, such that $n = N/V$.

  Figure \ref{fig:loading rate} shows the loading rate $R$ and steady-state atom number $N_{\rm{S}}$ as a function of Li beam temperature $T$.  We observe loading rates as high as \mbox{$5\cdot 10^8$\,s$^{-1}$}, comparable to that of many Zeeman slowers \cite{PeixotoThesis}.
The steady state atom number initially increases as $T$ is increased, reaching a maximum $N_{\rm{S}} \approx 10^7$.  For temperatures in excess of 330\,$^\circ$C, the increase in background pressure due to outgassing exceeds the increase in Li production, and the trapped atom number decreases marginally.  At the optimal $T=330\,^\circ$C, we measure an equilibrium N$_2$-equivalent pressure of $1.8\,(4)\cdot10^{-7}$\,Pa on an ion gauge attached to the vacuum chamber.

AMDs have proven useful for a variety of cold atom experiments.  The 3D-printed AMD presented here loads a Li MOT with comparable atom number, load rate, and background pressure to other MOTs directly loaded from effusive sources \cite{Moore2005,Muhammad2008,Ladouceur2009,Scherer2012}.  Reference \cite{Gunton2013} shows that with careful preparation of the Li, similar loading rates to those demonstrated here are attainable at operating pressures an order of magnitude lower (and thus $N_{\rm{S}}$ an order of magnitude higher).  We plan to use this 3D-printed AMD in designing a cold atom vacuum standard based on Li \cite{Scherschligt2017,Jousten2017}.  Our design is generic to a number of atomic species; specifically  Sr, Yb, Mg, and Ca all have negligible vapor pressure at room temperature, but can achieve vapor pressures comparable to Li at comparable ($\pm 50$\,K) operating temperatures.  Thus, this source warrants consideration for other deployable technologies based on cold atoms, including clocks and accelerometers.

\begin{acknowledgments}
The authors  thank M. Sefa for initial measurements of 3D-printed titanium outgassing and S. Maxwell and W. McGehee for useful comments on the manuscript.
\end{acknowledgments}
\bibliography{thebib}

\begin{thebibliography}{20}
\bibitem [{\citenamefont {{Any mention of commercial products within this work
  is for information only; it does not imply recommendation or endorsement by
  NIST}}()}]{NISTDisclaimer}%
  \BibitemOpen
  \bibfield  {author} {\bibinfo {author} {\bibnamefont {{Any mention of
  commercial products within this work is for information only; it does not
  imply recommendation or endorsement by NIST}}},\ }\href@noop {}
  {}\BibitemShut {NoStop}%
\end{thebibliography}

\clearpage

\end{document}